\documentclass[aps,prb,reprint,superscriptaddress]{revtex4-2}
\usepackage[dvipdfmx]{graphicx}
\usepackage{amsmath,mathcomp}
\usepackage{physics}
\usepackage{color}
\usepackage{ulem}

\usepackage{txfonts}

\begin{document}


\title{Electronic Structure of a Delafossite Oxide CuAlO$_2$ in Comparison with CuCrO$_2$
}

\author{Kenta Takahashi}
\affiliation{Department of Applied Physics, Tokyo University of Science, Katsushika, Tokyo 125-8585, Japan}

\author{Ryo Kato}
\affiliation{Department of Applied Physics, Tokyo University of Science, Katsushika, Tokyo 125-8585, Japan}

\author{Mario Okawa}\email[]{m-okawa@rs.tus.ac.jp}
\affiliation{Department of Applied Physics, Tokyo University of Science, Katsushika, Tokyo 125-8585, Japan}

\author{Tetsuji Okuda}
\affiliation{Department of Electrical and Electronics Engineering, Kagoshima University, Kagoshima 890-0065, Japan}

\author{Akira Yasui}
\affiliation{Japan Synchrotron Radiation Research Institute, Sayo, Hyogo 679-5198, Japan}

\author{Eiji Ikenaga}
\affiliation{Institute of Materials and Systems for Sustainability, Nagoya University, Nagoya 464-8601, Japan}

\author{Kanta Ono}
\affiliation{Photon Factory, KEK, Tsukuba, Ibaraki 305-0801, Japan}

\author{Noriaki Hamada}
\affiliation{Department of Physics, Tokyo University of Science, Noda, Chiba 278-8510, Japan}
\affiliation{Center for Spintronics Research Network, Osaka University, Toyonaka, Osaka 560-8531, Japan}

\author{Tomohiko Saitoh}\email[]{t-saitoh@rs.tus.ac.jp}
\affiliation{Department of Applied Physics, Tokyo University of Science, Katsushika, Tokyo 125-8585, Japan}

\date{\today}

\begin{abstract}
We investigated the electronic structure of a delafossite oxide CuAlO$_2$
in comparison with CuCrO$_2$, using vacuum ultraviolet (VUV) and hard x-ray (HX) photoemission spectroscopy (PES) and band structure calculations within the local-density approximation +$U$ scheme.
Cu 3$p\! -\! 3d$ resonant PES and Cu $2p$ core-level HX-PES give concrete evidence that the Cu valence in CuAlO$_2$ is monovalent, which makes a sharp contrast to CuCrO$_2$ that showed contradictory results depending upon probing methods [Yokobori \textit{et al.}, Phys. Rev. B \textbf{87}, 195124 (2013)].
We further noticed weak but detectable satellite structures due to Cu$^{2+}$ state in the Cu $2p$ HX-PES spectrum of CuCrO$_2$, surprisingly, without an observable main peak due to the Cu$^{2+}$ state.
Comparisons between the valence-band HX-PES spectra and the band structure calculations revealed a significant amount of the Cu $4s$ and Al $3s$/Cr $4s$ states in the valence band.
The above results can reasonably be explained by
the Cu $3d, 4s-$Cr $3d$ charge transfer model that we previously proposed in the above paper,
thereby demonstrating the critical role of empty $3d$ orbitals in the electronic structure of the copper delafossite oxides.
\end{abstract}


\maketitle


\section{\label{sec:Int}Introduction}

Delaffosite-type oxides Cu$M$O$_2$ ($M$= trivalent cation) have recently been attracting much interest because of their enormous potential for various applications,\cite{Daou17} such as a \textit{p}-type transparent conducting oxide (TCO) (CuAlO$_2$),\cite{Kawazoe97} a thermoelectric material (CuCr$_{1-x}$Mg$_x$O$_2$),\cite{Okuda05}
a multiferroic oxide (CuFeO$_2$ and CuCrO$_2$),\cite{Kimura06,Seki08}
or a photoevolution catalysit (CuMnO$_2$).\cite{Bessekhouad03}
Among them, Kawazoe \textit{et al.} reported CuAlO$_2$ as a first promissing  \textit{p}-type TCO with both high carrier mobility ($\sim$10 cm$^2$V$^{-1}$s$^{-1}$) and a wide band gap ($\sim$3.5 eV).\cite{Kawazoe97}
According to their strategy for designing a \textit{p}-type TCO, this was accomplished by hole doping into a wide gap Cu$^+$ oxide with the $d^{10}$ closed-shell electron configuration.\cite{Kawazoe97} Hence the top of the valence band was expected to have the Cu $3d$ character with some O $2p$ character through the hybridization.

Following this strategy, other Cu$M$O$_2$ such as CuCrO$_2$ have been investigated in the contex of TCO.\cite{Scanlon09,Arnold09, Hiraga11} However, there were incompatible reports on the electronic structure of CuCrO$_2$ whether the Cr $3d$ or Cu $3d$ character was dominant at that point.\cite{Scanlon09,Arnold09, Hiraga11,Maignan09}
Yokobori \textit{et al.} has put an end of this controversy by showing the direct evidence of the dominant Cr $3d$ character near the Fermi level ($E_{\text{F}}$) in photoemission spectra.\cite{Yokobori13}
At the same time, however, they have also discovered a curious set of valence and core-level photoemission spectra and a Cu $L$ edge soft x-ray absorption spectrum which did not seem to be consistent with each other in terms of the Cu valence in this compound:
They observed a clear satellite structure of Cu$^{2+}$ in their resonant photoemission spectra while they failed to observe spectral weight of Cu$^{2+}$ in their Cu $2p$ core-level and Cu $L$ edge soft x-ray absorption spectra.\cite{Yokobori13}
In order to explain their experiment, they have proposed a unique but rather complicated model for the electronic structure of CuCrO$_2$.\cite{Yokobori13} The essential point of the model is an appreciable charge transfer from the Cu $4s$ state to the Cr $3d$ state, which can only be realized in the presense of available empty $M$ states near $E_{\text{F}}$ in Cu$M$O$_2$.

In order to confirm whether this model is appropriate or not, CuAlO$_2$ is the most suitable compound because Al is a typical metal and
thus the valence band of CuAlO$_2$ consists of Cu $3d$ (and $4s$) and O $2p$ states with no nominal electronic states from Al atoms in the valence band. 
As mentioned above, the electronic structure of CuAlO$_2$ has been investigated for decades mostly in terms of \textit{p}-type TCO. 
While so many theoretical studies with band structure calculations have been done,\cite{Yanagi00,Robertson02,Koyanagi03,Falabretti07,Scanlon09,Gillen11,Nakanishi12,Xu12,Mori13,Thatribud14,Wang17,Liu17}
only few photoemission studies have been reported so far.\cite{Yanagi00,Aston05,Arnold09,Liu17}

Yanagi \textit{et al.} performed x-ray photoemission spectroscopy (XPS), ultraviolet photoemission spectroscopy (UPS), and inverse photoemission spectroscopy (IPES). Comparing with their band strcuture calculations, they concluded that the top of the valence band was dominated by the Cu-O bonding states with negligible  contribution from Al $3p$ states. They also estimated the size of the energy gap as about 3.5 eV, which was reasonably consistent with the direct optical gap.\cite{Yanagi00}
Arnold \textit{et al.} compared the valence band electronic structure of CuCrO$_2$ and CuAlO$_2$ using XPS.\cite{Arnold09} They observed a satellite structure at $\sim$15.5 eV in both compound and interpreted as the Cu $3d^84s$ final state, which typically appears in a Cu$_2$O spectrum at 15.3 eV.
This satellite structure is a key to determine the Cu valence because it is a direct evidence for Cu$^+$. In connection with this, the 13-eV satellite observed in CuO at 12.5$-$12.9 eV is also important as a direct evidence for Cu$^{2+}$.\cite{Thuler82,Ghijsen90,Shen90}
Because the intensity of these satellites is enhanced in Cu $3p\! -\! 3d$ or $2p\! -\! 3d$ resonant photoemission spectroscopy, Yokobori \textit{et al.} made use of the two satellites to determine the Cu valence in CuCrO$_2$.\cite{Yokobori13}

In this paper, we investigate the electronic structure of CuAlO$_2$ using $3p\! -\! 3d$ resonant photoemission spectroscopy, hard x-ray photoemission spectroscopy (HX-PES), and band structure calculations using the local-density approximation (LDA)+$U$ method. For comparison, we also report the Cu $2p$ and the valence-band HX-PES spectra of CuCrO$_2$.


\section{\label{sec:Exp}Experiment and Calculation}

Polycrystalline samples of CuAlO$_2$ were prepared by the standard solid-state reaction.\cite{Okuda05} 
Vacuum ultraviolet (VUV)-PES measurements in the range of the Cu 3$p\! -\! 3d$ resonance ($h\nu$=70$-$85 eV) were performed at BL-28A of the Photon Factory, KEK, using a SCIENTA SES-2002 electron analyzer. HX-PES spectra taken with $h\nu$=7940 eV were measured at BL09XU and BL47XU of SPring-8 using SCIENTA R4000 electron analyzers.
In order to obtain clean surface, we fractured the samples $in$ $situ$ right before the measurements. The fracturings and the measurements were done in ultrahigh vacuum, namely, about 2.0$\times$10$^{-8}$ Pa (VUV-PES), about 1.2$\times$10$^{-7}$ Pa (fracturing for HX-PES), and about 5$\times$10$^{-7}$ Pa (measurment for HX-PES), all at 300 K.  The intensity of the resonant PES spectra was normalized using photon current of the exit mirror.
The energy resolutions were 30 meV (VUV-PES) and 250 meV (HX-PES) FWHM. All the Fermi level ($E_{\text{F}}$) positons in the experiments were calibrated with Au spectra.
In the polarization-dependent HX-PES measurements, diamond phase retarders were used to switch the horizontally and vertically linear-polarized light.

We also performed band structure calculations 
using the code ABCAP
with the full potential linearized augmented plane-wave (FLAPW) method \cite{Andersen75,Takeda79} in the LDA+$U$ scheme.\cite{Hohenberg64,Kohn65,Vosko80,Anisimov91,Solovyev96,Anisimov97}
We adopted 2.0 eV of the effective Coulomb repulsion $U_{\rm{eff}}=U-J$ at the Cu site with $J=0.7$ eV.
The rhombic lattice parameters ($a$=2.858 \AA, $c$=16.958 \AA) were taken from Ref.~\onlinecite{Hamada_I06}. The plane-wave cut off energy was 163 eV for the wave function. We took 72 $k$ points in the irreducible Brillouin zone for the rhombohedral Brillouin zone.

\section{\label{sec:Results}Results}

\subsection{$3p\! -\! 3d$ resonant photoemission spectra in the Cu charge-transfer satellite region}

\begin{figure}[h]
	\begin{center}
	\includegraphics[width=82mm,keepaspectratio]{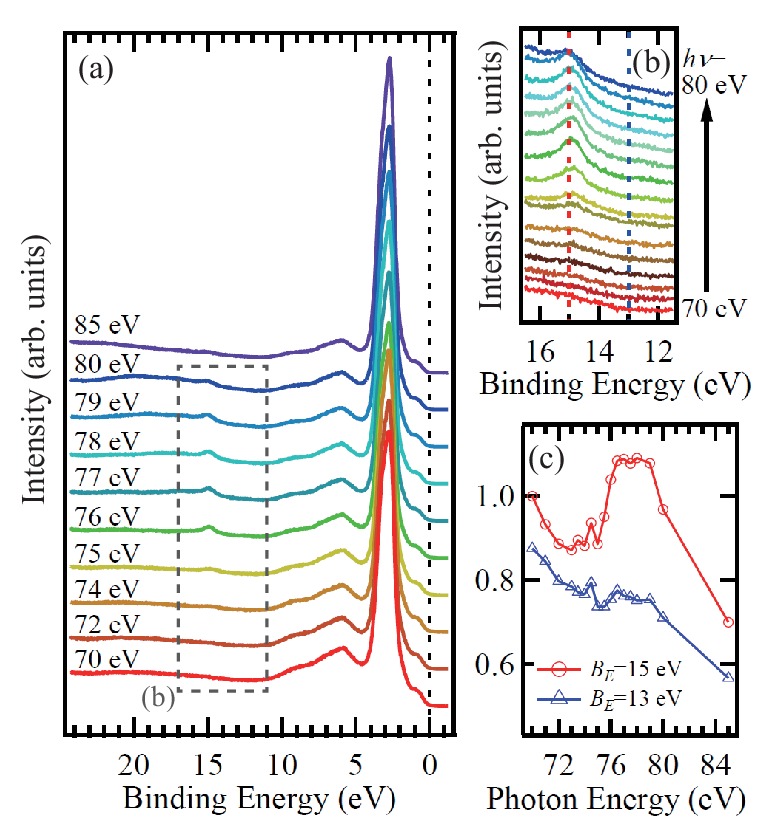}
	  \caption{(Color oneline) (a) Valence-band spectra of CuAlO$_2$ in the photon energy range of the Cu $3p\! -\! 3d$ resonance. (b) Detailed satellite structure of (a) in an expanded energy scale. (c) CIS spectra at the binding energy ($E_{\rm{B}}$) of 13 and 15 eV, corresponding to the Cu$^{2+}$ and Cu$^{+}$ satellite, respectively.}
\label{FIG_VUV-Sat}
\end{center}
\end{figure}

It is well-established that the valence-band photoemission spectra of CuO and Cu$_2$O show characteristic satellite structures at 12.5$-$12.9 eV and 15.3 eV, which have mainly $3d^8$ and $3d^84s$ final-state character, respectively.\cite{Thuler82,Ghijsen90,Shen90}
Moreover, the two satellites have different resonant photon energies, 74 and 77 eV, respectively, in the range of Cu $3p\! -\! 3d$ resonance.
Therefore, the  Cu $3p\! -\! 3d$ resonant photoemisson spectroscopy is a direct probe to identify the Cu valence in Cu compounds.


Figure~\ref{FIG_VUV-Sat}~(a) shows the valence-band photoemission spectra of CuAlO$_2$ taken across the resonant energies of the Cu satellite structures. 
The valence band has a sharp and intensce main peak at $\sim$3 eV with a shoulder structure in the lower binding energy side of the main peak. One can observe that a structure around 6 eV is not a shoulder of but a separated feature from the main peak, which is a characteristic difference between Cu$_2$O and CuO.\cite{Thuler82,Ghijsen90}
All these observations demonstrates that the valence band of CuAlO$_2$ is quite similar to that of Cu$_2$O.
Figure~\ref{FIG_VUV-Sat}~(b) shows satellite region of Fig.~\ref{FIG_VUV-Sat}~(a) in an expanded scale. It is obvious that the 15-eV satellite shows the resonant enhancement at the photon energy of $76\! -\! 79$ eV while no ditectable enhancement can be seen in the 13-eV satellite from 70 eV through 80 eV.

The constant initial state (CIS) spectra of these two satellite peaks are shown in Fig.~\ref{FIG_VUV-Sat}~(c). The CIS profile of the 15-eV satellite is almost identical to that of Cu$_2$O.\cite{Thuler82,Ghijsen90}
The rather broad plateau structure is due to the 3$p_{3/2}$ and 3$p_{1/2}$ spin-orbit splitting.
On the other hand, the CIS profile of the 13-eV satellite has no appreciable enhancement as observed in Fig.~\ref{FIG_VUV-Sat}~(b).
%

The result of Fig.~\ref{FIG_VUV-Sat} makes a sharp contrast to the case of CuCrO$_2$, in which two distinct satellite structures at 13 eV and 15 eV were observed.\cite{Yokobori13}
In fact, this difference was  found in the  Al $K\alpha$ XPS spectra of CuAlO$_2$ and CuCrO$_2$ by Arnold \textit{et al.};\cite{Arnold09}
They already pointed out that the 15-eV satellite in both compounds originated from the $3d^84s$ final-state character, whereas they made no comment on the 13-eV satellite region.
In their paper, one can notice that their CuCrO$_2$ spectrum has some intensity around 13 eV while the CuAlO$_2$ one has a clear dip there.
This difference completely agrees with what were observed in our present and previous studies.

\subsection{Cu $2p$ core level}

\begin{figure}[h]
	\begin{center}
\includegraphics[width=85mm,keepaspectratio]{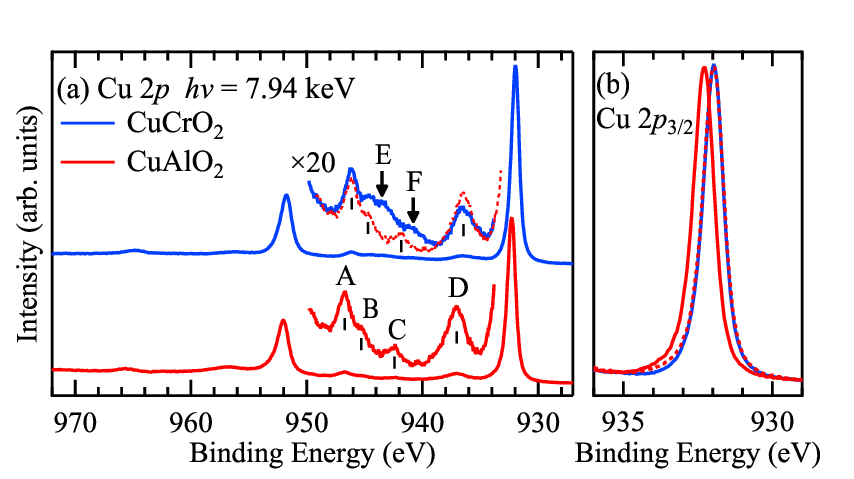}
\caption{(Color online) (a) Cu $2p$ core-level HX-PES spectra of CuAlO$_2$ (red line) compared with CuCrO$_2$ (blue line)\cite{Yokobori13} with the expanded ones in the satellite region. Two arrows denote the intensity due to the charge-transfer satellite of the Cu$^{2+}$ initial state, which is observable only in CuCrO$_2$. Note that the expanded CuAlO$_2$ spectrum superimposed on the CuCrO$_2$ one (red dotted line) is shifted by -0.59 eV.
(b) Detailed comparison of the $2p_{3/2}$ peak of CuAlO$_2$ and CuCrO$_2$. Red dotted line denotes the CuAlO$_2$ spectrum with a shift of -0.32 eV.}
\label{FIG_Cu2p}
\end{center}
\end{figure}

Figure~\ref{FIG_Cu2p} compares Cu $2p$ core-level HX-PES spectra of CuAlO$_2$ and CuCrO$_2$ \cite{Yokobori13}.
The whole Cu $2p$ spectra of the two compounds in Panel (a) look almost identical; both show the obvious Cu$^{+}$ features of the Cu $2p$ core-level spectrum, namely the sharp Cu $2p_{3/2}$ and $2p_{1/2}$ peaks 
and very weak satellite structures typically observed in the CuO spectrum \cite{Vasquez98-Cu2O,Li-Junqiang14}.
Panel (a) also agrees well with the reported conventional XPS spectra of CuAlO$_2$ \cite{Aston05} and CuCrO$_2$ \cite{Arnold09,Le11}, respectively.
However, we can find a subtle but significant difference between the two spectra in the satellite region; 
Structures A, B, C, and D in the CuAlO$_2$ spectrum are the Cu$^{+}$ satellite structures, which are identical to s1, s1', s2, and m2 in the CuAlO$_2$ spectrum by Aston \textit{et al.}\cite{Aston05,Note4}
These four satellites are all observed also in the CuCrO$_2$ spectrum. The non-negligible difference is that the CuCrO$_2$ spectrum has extra broad features E and F, whose binding energy indeed matches that of the Cu$^{2+}$ satellites in oxidized CuAl$_{1-x}$Zn$_x$O$_2$ or CuRh$_{1-x}$Mg$_x$O$_2$\cite{Aston05,Le11,Note4}.
Nevertheless, Panel (b) clearly demonstrates that the lineshapes of the Cu $2p_{3/2}$ main peak of the two compounds are almost identical with no detectable signature of the Cu$^{2+}$ main peak, which should be observed at about 934 eV.\cite{Aston05,Le11,Vasquez98-CuO}
In other words, the Cu$^{2+}$ component in CuCrO$_2$, if exists, seems to have the satellite intensity alone without the main peak intensity.

Such an unusuall behavior should be a manifestation of the complex electronic structure in CuCrO$_2$ and be understood in the context of the Cu-O-Cr charge-transfer model that we previously proposed\cite{Yokobori13}. We will discuss this issue later in Sec.~\ref{sec:Discuss}.

It should also be noted that the binding energy difference between the main peak and the Cu$^{+}$ satellites is smaller in CuAlO$_2$ than CuCrO$_2$ by $\sim$0.27 eV.
This would be reflecting a smaller effective hybridization between Cu $3d$ or $4s$ and O $2p$ states in CuAlO$_2$, indicating some role of the Cr $3d$ states (and probably $4s$ states as well).

\subsection{Band structure calculations}

\begin{figure}[h]
	\begin{center}
 	\includegraphics[width=80mm,keepaspectratio]{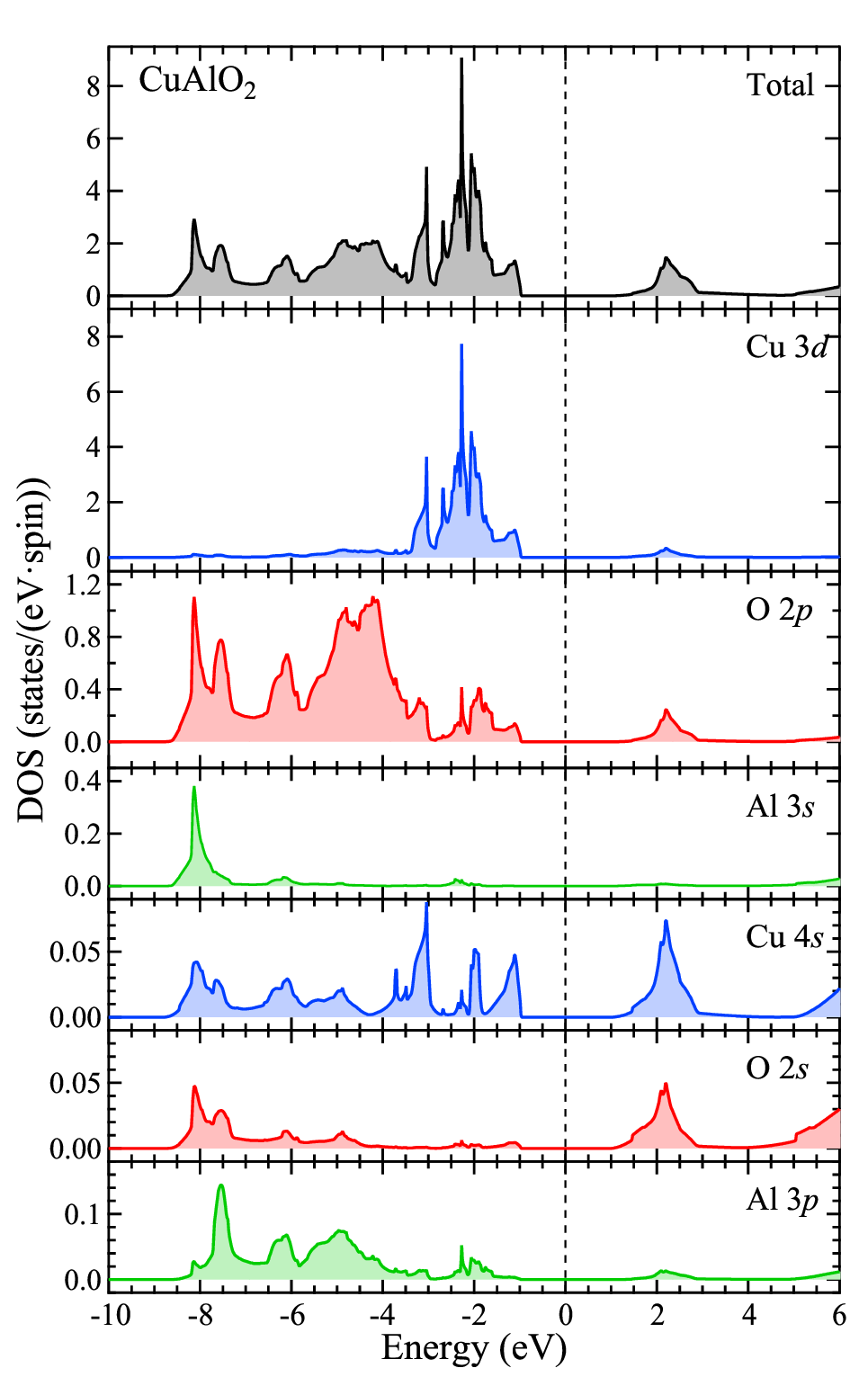}
 	\caption{(Color online) LDA+$U$ band structure calculations of CuAlO$_2$. $U_{\rm{eff}}$ was set to 2 eV for Cu $d$ states.}
\label{FIG_LDAU}
\end{center}
\end{figure}

\begin{figure}[h]
\begin{center}
\includegraphics[width=75mm,keepaspectratio]{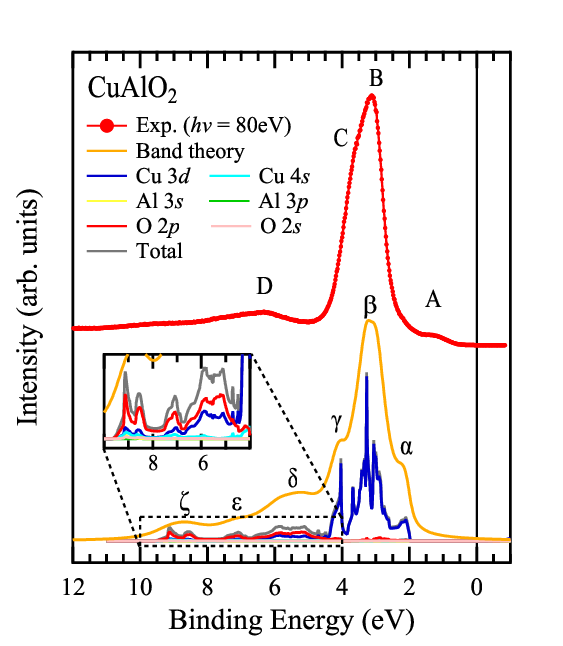}
\caption{(Color online) Comparison between the calculated DOS and the valence-band spectrum taken at $h\nu$=80.0 eV. 
Inset shows the expanded figure of the theory between 4 and 10 eV.
The theory is shifted by 1.0 eV in the binding energy axis from the original calculation so that the major peak ($\beta$) matches the experimental one (B). Note that this shift is smaller than the calculated bandgap.
}
\label{FIG_LDAU-UPS}
\end{center}
\end{figure}

To better understand the valence-band electronic structure, we performed LDA+$U$ band structure calculations.
Figure~\ref{FIG_LDAU} shows the result of our LDA+$U$ calculations.
The calculated DOS is consistent with the literature \cite{Yanagi00,Falabretti07,Gillen11}.
Because all the Al $3s$ and $3p$ orbitals are nominally empty, the valence band mostly consists of Cu $3d$ and O $2p$ states.
The Cu $3d$ partial DOS has intense peaks between 1.0 and 3.5 eV whereas
the DOS from 3.5 to 8.5 eV (the bottom of the valence band) is mostly due to the O $2p$ partial DOS.
It is noted that the nominally empty Al $3s$ and $3p$ states actually have small but non-negligible DOS below 0 eV, particularly in the two peaks at the bottom of the valence band. Compared with them, the Cu $4s$ and the O $2s$ DOS are quite negligible in the figure.

A comparison between the spectrum taken at $h\nu=80$ eV and the calculation is shown in Fig.~\ref{FIG_LDAU-UPS}.
The theoretical simulation curve has been constructed by broadening the cross-section-weighed sum of the Cu $3d$, $4s$, Al $3s$, $3p$, and O $2p$, $2s$ partial DOS with an energy dependent Lorentzian function due to the lifetime effect and a Gaussian due to the experimental resolution.\cite{Saitoh97,Iwasawa09,Yeh85,Note1}

The theoretical spectrum shows that the leading shoulder structure at the top of the valence band ($\alpha$), the major peak ($\beta$), and the shoulder structure ($\gamma$) are dominated by the Cu $3d$ states
whereas the following structures ($\delta, \varepsilon, \zeta$) primarily due to the O $2p$ states with a minor contribution of the Cu $3d$ states. The intensity of the other states are very small because of their small photoionization cross sections.\cite{Yeh85}

One can see that the theoretical spectrum basically reproduces the experimental one; the experimental structures A to D can be assigned to the theoretical structures $\alpha$ to $\delta$, respectively. There are, however, some differences. For example, the location of A is rather closer to $E_{\text{F}}$ than $\alpha$ and the intensity of A and D is considerably smaller than theory.

\subsection{Valence-band spectra with hard x-ray photoemission}

\begin{figure}[htb]
\begin{center}
\includegraphics[width=70mm,keepaspectratio]{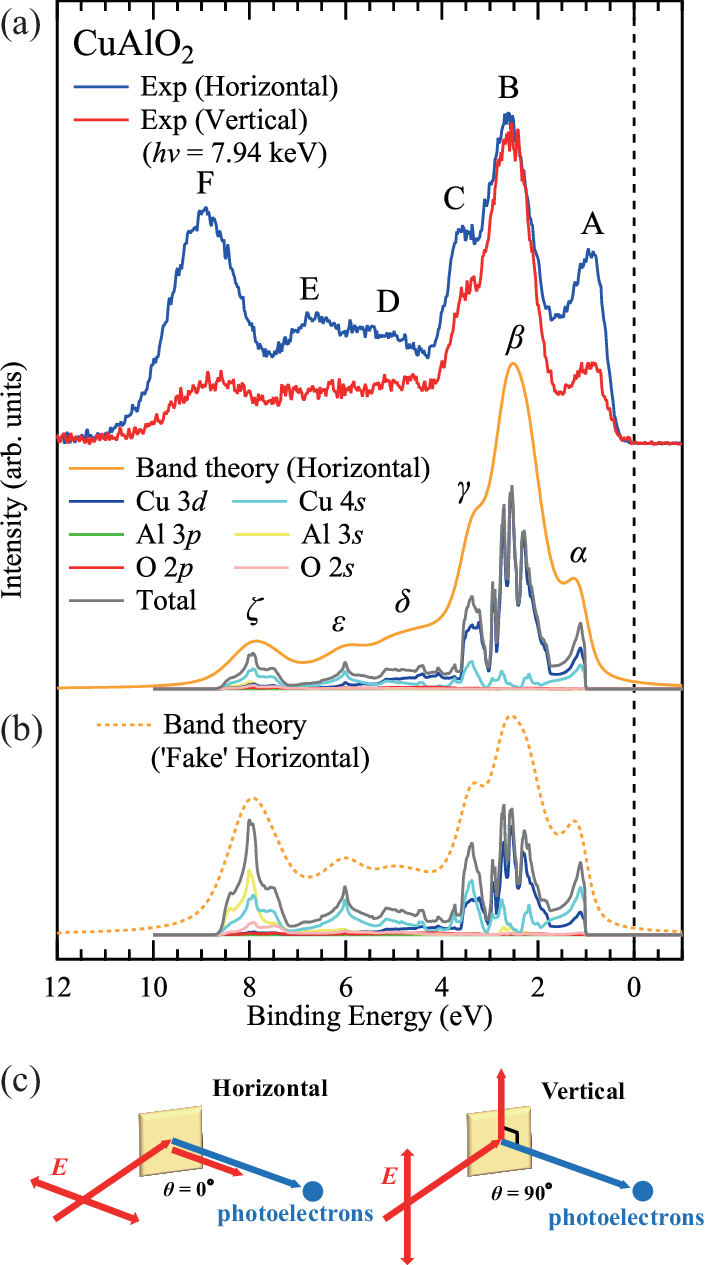}
\caption{(Color online)
(a) Valence-band HX-PES spectra taken at $h\nu$=7.94 keV in two experimental geometries compared with
the simulation curve constructed from the band theory. The cross-section-weighted partial DOS's are also shown.
(b) A band-theory simulation curve with overestimated partial DOS of the $s$
states. See the text for more detail.
(c) Illustration of the two experimental geometries.
}
\label{FIG_LDAU-HXPES}
\end{center}
\end{figure}

Figure~\ref{FIG_LDAU-HXPES} shows HX-PES valence-band spectra taken at $h\nu=$7.94 keV compared with the band theory.
Here, we performed the experiment in two geometroies using linearly polarized hard x-ray in order to see how and to what extent the Cu $4s(p)$ and Al $3s(p)$ states contribute to the valence-band electronic structure.
Figure~\ref{FIG_LDAU-HXPES} (c) shows the two experimental geometories; in the ``Horizontal" geometry, which is a typical geometry in experiments at beamlines, the polarization vector of the incident light and the direction of the detected photoelectrons are paralell, while in the ``Vertical" geometry, they are perpendicular to each other.
It is known that 
the theoretical and experimental photoionization cross section of $s$ and $np\ (n\! >\! 4)$ states are strongly suppressed in the Vertical geometry, whereas that of $d$, $f$, and $np\ (n\! \le\! 4)$ have relatively weak polarization dependence
\cite{Sekiyama10,Weinen15,Trzhaskovskaya01,Trzhaskovskaya06}.
Hence, in our experiment,
the spectral weight of the Cu $4s$ and the Al $3s$ states in the Horizontal geometry is rather enhanced while that in the Vertical geometry is strongly suppressed and must be negligible.
On the other hand, it is also known that the photoionization cross section of 
$s$ and $p$ states is comparable to that of $d$ or $f$ states in the hard x-ray region \cite{Yeh85,Trzhaskovskaya01,Trzhaskovskaya06}.
Therefore, if there exit considerable contributions of  the Cu $4s$ and/or the Al $3s$ states in the valence band, the two experimental spectra must show significant differences.

Figure~\ref{FIG_LDAU-HXPES} (a) compares the experimental spectra in the two geometries 
and the band theory that simulates the spectrum in the Horizontal geometry.\cite{Note2}
The simulation curve was constructed in the same way as in Fig.~\ref{FIG_LDAU-UPS}, including the effects of the Horizontal geometry \cite{Trzhaskovskaya01,Trzhaskovskaya06}.
The figure demonstrates that the experimental lineshape of the valence band largely changes with the polarization direction and thus the $s$ states indeed significantly contribute to the whole valence band, particularly at the top and the bottom of the valence band. 

One can identify the experimental structures A to F in the two spectra as the theoretical ones $\alpha$ to $\zeta$.
Nevertheless, the simulation curve for the Horizontal geometry fails to reproduce the spectrum in the Horizontal geometry but rather reproduces well that in the Vertical geometry.
The disagreement is most likely because a large portion of the $s$ states (and the $p$ states, too) exists out of the muffin-tin radius, thereby being not counted in the partial DOS.
This idea can be confirmed by constructing a `fake' Horizontal band theory curve in which the contribution of the $s$ states is intentionally overestimated.
Figure~\ref{FIG_LDAU-HXPES} (b) shows such a simulation curve. Here, the
photoionization cross sections of the Cu $4s$, Al $3s$, and O $2s$ are 120, 90, and 30 times larger than the correct values, respectively. Although these numbers have no physical meaning, we find that this `fake' Horizontal curve well reproduces the Horizontal experiment.

Both the experiment and the theory thus show a significant amount of the
Cu $4s$ and Al $3s$ states in the valence band of CuAlO$_2$.
However, the factor of $\sim$100 for the theoretical cross section in the `fake' Horizontal simulation may be too large even though the much portion of the $s$ states are lost in the partial DOS.
Hence, further investigations such as re-estimation of the $s$ partial DOS would be required for detailed comparisons between the polarization-dependent valence-band HX-PES spectra and band structure calculations.

\section{\label{sec:Discuss}Discussion}

In our previous study, we proposed that the electronic structure of CuCrO$_2$ was able to be described by finite Cu $3d,4s-$Cr $3d$ charge transfer via O $2p$ states in the ground-state \cite{Yokobori13}.
This model can explain many of the physical properties of CuCrO$_2$ and also predicts that the valence-band photoemission spectrum of CuAlO$_2$ should not have the Cu$^{2+}$ satellite structure at $\sim$13 eV because no available empty state exists at Al site. 

We believe that the Cu-O-Cr charge transfer model is now established in the present study:
Firstly, the Cu $3p$$-$$3d$ resonant photoemission spectra of CuAlO$_2$
directly confirms the model.
Secondly, the fact that the weak but finite Cu$^{2+}$ satellite structure in the Cu $2p_{3/2}$ core level was observed in CuCrO$_2$ but not in CuAlO$_2$ should be interpreted as a result of a small amount of charge transfer from the Cu site to the O and/or Cr sites in CuCrO$_2$.

We also point out that this unusual Cu$^{2+}$ satellite without detectable intensity of the main peak would be explained within the model as follows:
The model \cite{Yokobori13} describes the ground state of the Cu-O-Cr cluster in CuCrO$_2$ (CCO in short) $|g\rangle_{\rm{CCO}}$ as
\begin{equation*}
|g\rangle_{\rm{CCO}} =
\alpha|d^{10} d^3 \rangle +
\beta|d^{9}\! s\ d^3 \rangle +
\gamma|d^{9} d^4 \rangle +
\delta|d^{10} \underline{L}\ d^4 \rangle ,
\end{equation*}
where the left $d^9$ and $d^{10}$ denote the Cu $3d$ states, $s$ denotes the Cu $4s$ state, and the right $d^3$ and $d^4$ denote the Cr $3d$ states. 
$|d^{10} d^3 \rangle$ is the leading major configuration and
$|d^{9}\! s\ d^3 \rangle$ corresponds to the $d_{3z^2-r^2}$ hole state
\cite{Zuo99}. They consist of the Cu$^{+}$ electronic configuration like in Cu$_2$O and CuAlO$_2$ as well.
On the other hand, $|d^{9} d^4\rangle$ and $|d^{10} \underline{L}\ d^4 \rangle$ originate from 
the Cu $4s$-to-Cr $3d$ charge transfer and
the O $2p$-to-Cr $3d$ charge transfer, respectively, both corresponding to the Cu$^{2+}$ component.

Then, the final state of the Cu $2p$ core-level photoemission can be described as
\begin{equation*}
| f^{\rm{Cu}}_c \rangle_{\rm{CCO}} =
a|\underline{c}d^{10} d^3 \rangle +
b|\underline{c}d^{9}\! s\ d^3 \rangle +
c|\underline{c}d^{10} \underline{L}\ d^4 \rangle +
d|\underline{c}d^{9} d^4 \rangle ,
\end{equation*}
where $\underline{c}$ denotes a Cu $2p$ core hole.
In this formula, 
$|\underline{c}d^{10} d^3 \rangle$ and
$|\underline{c}d^{9}\! s\ d^3 \rangle$ describe the Cu $2p_{3/2}$ main peak and its weak satellites of the Cu$^{+}$ state, respectively, while
$|\underline{c}d^{10} \underline{L}\, d^4 \rangle$ and
$|\underline{c}d^{9} d^4 \rangle$ should roughly correspond to
the main peak and its large satellites of the Cu$^{2+}$ state, respectively.

In $| f^{\rm{Cu}}_c \rangle_{\rm{CCO}}$,
the weight of 
$|\underline{c}d^{10} \underline{L}\, d^4 \rangle$ and
$|\underline{c}d^{9} d^4 \rangle$ should be very small
because the strong screening effects by
the positive Cu $2p$ core hole prompt the process
$|\underline{c}d^{10} \underline{L}\, d^4 \rangle \to
|\underline{c}d^{10} d^3 \rangle$
and the process
$|\underline{c}d^{9} d^4 \rangle \to
|\underline{c}d^{9}\! s\ \underline{L}\ d^4 \rangle \to
|\underline{c}d^{9}\! s\ d^3 \rangle$,
both due to the 
$\underline{c}-\underline{L}$ Coulomb repulsion.
We already pointed out that this can explain that the Cu $2p$ spectrum of CuCrO$_2$ is almost Cu$^{+}$ like \cite{Yokobori13}. 
In addition, here, we further infer that the screening effects would be larger in $|\underline{c}d^{10} \underline{L}\, d^4 \rangle$ than $|\underline{c}d^{9} d^4 \rangle$
simply because
the latter has two successive processes,
thereby leading the weak but finite Cu$^{2+}$ satellites with the negligibly small Cu$^{2+}$ main peak as displayed in Fig.~\ref{FIG_Cu2p}.
To summarize, the final state of the Cu $2p$ core-level photoemission of CuCrO$_2$ and CuAlO$_2$ (CAO in short) can be described as
%
\begin{align*}
| f^{\rm{Cu}}_c \rangle_{\rm{COO}} 
&=
a|\underline{c}d^{10} d^3 \rangle +
b|\underline{c}d^{9}\! s\, d^3 \rangle +
d|\underline{c}d^{9} d^4 \rangle \\
| f^{\rm{Cu}}_c \rangle_{\rm{CAO}} 
&=
a'|\underline{c}d^{10} \rangle +
b'|\underline{c}d^{9}\! s \rangle .
\end{align*}
%

\begin{figure}[htb]
\begin{center}
\includegraphics[width=70mm,keepaspectratio]{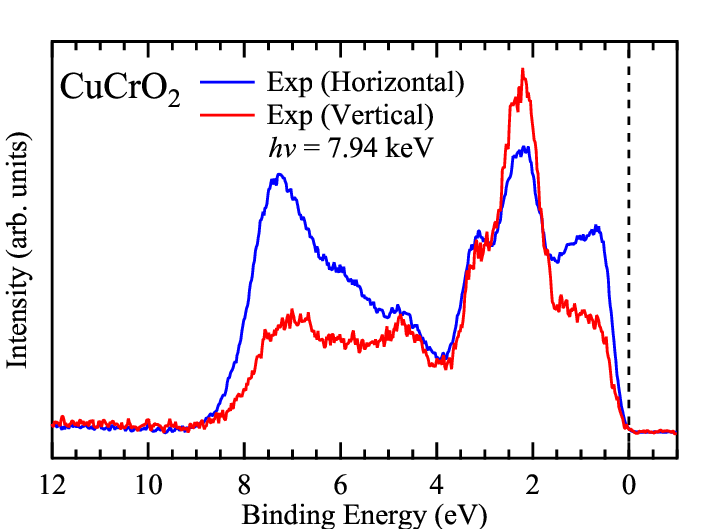}
\caption{(Color online)
Valence-band HX-PES spectra of CuCrO$_2$ taken at $h\nu$=7.94 keV in the two experimental geometries.
}
\label{FIG_CuCrO2-HXPES}
\end{center}
\end{figure}

In our model, a meaningful amount of the Cu $4s$ states and also the hybridizing Cr $4s$/Al $3s$ states should be important.
The polarization-dependent study clearly demonstrates the significant spectral weight of the $s$ states particularly at the top and the bottom of the valence band not only in CuAlO$_2$ (Fig.~\ref{FIG_LDAU-HXPES}), but also in CuCrO$_2$ as shown in Fig.~\ref{FIG_CuCrO2-HXPES},\cite{Note3}
strongly supporting our charge-transfer model. 

Moreover, Figs.~\ref{FIG_LDAU-HXPES} and \ref{FIG_CuCrO2-HXPES} imply that such an amount of the $s$ states are rather common in the valence band of the copper delafossite oxides Cu$M$O$_2$.
Therefore, the critical difference between CuAlO$_2$ (Cu$^{+}$) and CuCrO$_2$ (Cu$^{+}$ with some Cu$^{2+}$) is that the empty Cr $3d$ states near $E_{\rm{F}}$ are available for the charge-transfer process in CuCrO$_2$.
In fact, our model can explain the fact that the higher electrical conductivity of a hole-doped CuCrO$_2$ than other hole-doped Cu$M$O$_2$ with $M$ of typical element metals.\cite{Nagarajan01}
In this sense, the family of the copper delaffosite oxides Cu$M^{3+}$O$_2$ can be classified into two categories in terms of the valence-band electronic structure, $M$ is a transition-metal or a typical element metal.

\section{Conclusions}

We have studied the electronic structure of a delafossite oxide CuAlO$_2$ in comparison with CuCrO$_2$ with VUV photoemission spectroscopy, hard x-ray photoemission spectroscopy, and LDA+$U$ band-structure calculations.
The Cu $3p-3d$ resonant PES spectra and the Cu $2p$ HX-PES spectrum of CuAlO$_2$ established the monovalent Cu nature of this material, thereby confirming the Cu-O-Cr charge-transfer model that we have recently proposed \cite{Yokobori13}.
By contrast, the weak but detectable satellite structures due to Cu$^{2+}$ state were observed in the Cu $2p$ HX-PES spectrum of CuCrO$_2$. Interestingly, however, the Cu$^{2+}$ satellites did not accompany the main peak.
We point out that such an unusual result can also be explained by the Cu-O-Cr charge-transfer model.

The comparisons between the valence-band HX-PES spectra in the two experimental geometries and the LDA+$U$ band structure calculations revealed a siginificant amount of the Cu $4s$ and Al $3s$/Cr $4s$ states in the valence band, again confirming the Cu-O-Cr charge-transfer model.
Based on this model, we propose a classification of the copper delaffosite oxides Cu$M$O$_2$ into the two categories, one is that $M$ is a transition metal and the other is that $M$ is a typical element metal.

\begin{acknowledgments}

The authors would like to thank J. Izumi for her assistance in analyzing the data.
Constructions of X-ray optics in all the HX-PES beamlines are owing to collaborations with H. Ohashi, Y. Senba, T. Koyama and N. Kawamura of the Japan Synchrotron Radiation Research Institute.
This work was supported by JSPS KAKENHI Grant Numbers 25400378, 26400321, 16K06141, 17K05502, 18K03543, and 18K03550, and JST CREST Grant No. JPMJCR15Q2.
The synchrotron radiation experiments at the Photon Factory and SPring-8 were performed under the approval of the Photon Factory (2012G146, 2013G589, 2015G568, 2012S-001, and 2015S2-003) and of the Japan Synchrotron Radiation Research Institute (2014B1011, 2015A1008, 2015B1019, 2016A1013, and 2016A1142), respectively.

\end{acknowledgments}

%

\end{document}